\documentclass[floatfix,reprint,prl,aps,twocolumn,superscriptaddress,longbibliography]{revtex4-2}
\usepackage[T1]{fontenc}
\usepackage{graphicx}
\usepackage{enumitem}
\usepackage{amsmath}

\usepackage{amsfonts}
\usepackage{lipsum}

\usepackage{amsmath}
\usepackage{amsfonts}
\usepackage{amssymb}
\usepackage{mwe}
\usepackage{graphicx}
\usepackage{color}
\usepackage{bm}
\usepackage{soul}
\usepackage[normalem]{ulem}

\usepackage[colorlinks,bookmarks=false,urlcolor=blue, 
citecolor=blue, linkcolor = blue]{hyperref}

\begin{document}
\title{Itinerant magnetism and magnetic polarons in the 
triangular lattice Hubbard model}
\author{Ivan Morera}
\affiliation{Institute for Theoretical Physics, ETH Zurich,
Wolfgang-Pauli-Str. 27, 8093 Zurich, Switzerland}
\author{Eugene Demler}
\affiliation{Institute for Theoretical Physics, ETH Zurich,
Wolfgang-Pauli-Str. 27, 8093 Zurich, Switzerland}

\begin{abstract}
We use density matrix renormalization group to investigate the phase diagram of the Fermi Hubbard model on a triangular lattice with densities above half-filling, $1 \leq n < 2$. We discuss the important role of kinetic magnetism and magnetic polarons. For strong interactions and low doublon dopings, attractive interaction between polarons results in phase separation between the fully polarized state at finite doping and the commensurate spin-density wave state at half-filling. For intermediate interaction strength and small doping, competition between antiferromagnetic superexchange and kinetic magnetism gives rise to the incommensurate spin density wave (I-SDW) phase. Fully polarized ferromagnetic (FPF) phase for weak interactions is limited to dopings close to the van Hove singularity in the density of states. With increasing interactions the FPF phase expands to lower dopings. For strong interactions it reaches the low doping regime and is better understood as arising from proliferation of Nagaoka-type ferromagnetic polarons. Other phases that we find include a partially polarized phase, another type of I-SDW at high densities, and M\"uller-Hartmann ferromagnetism close to the band insulating regime. 
\end{abstract}
\maketitle

\textbf{Introduction.} 
Recent experimental studies of electrons in van der Waals moir\'e materials revealed unusual character of magnetism in triangular lattices. Experiments with heterobilayer transition metal dichalcogenide (TMDc) compounds indicated that magnetic correlations are absent in the Mott insulating phase at half-filling ($n=1$) but appear upon doping~\cite{Tang2020,ciorciaro2023kinetic,Tao2023}. This can be contrasted to square lattices, in which strongest magnetic order is observed in the Mott state at $n=1$, while doping tends to suppress it. These findings provide motivation for theoretical analysis of magnetic phases in the Hubbard model on a triangular lattice with finite doping (for theoretical studies of magnetism at filling $n=1$, see Refs.~\cite{Morita2002,Honerkamp2003,Shimizu2003,Koretsune2007,Sahebsara2008,Clay2008,Watanabe2008,Galanakis2009,Yoshioka2009,Yang2010,Kokalj2013,Yamada2014,Li2014,Tocchio2014,Laubach2015,Mishmash2015,Shirakawa2017,Szasz2020,Szasz2021,Cookmeyer2021,Wietek2021,Chen2022,Downey2023}). Prior work suggested several unusual phenomena upon doping the Mott insulator state~\cite{HS2005,Haerter2006,Haerter2006b,Sposeti2014,Batista2017,Morera2021,Lee2023,Morera2023,Davydova2023}. Delocalization of holes strengthens the 120$^{\circ}$ Neel order~\cite{Haerter2006,Zhu2022}, whereas doublons tend to favor ferromagnetic correlations in accordance with Nagaoka-Thouless theorem~\cite{Nagaoka1966,Thouless1965}. Other types of itinerant ferromagnetism are also expected to be relevant in the triangular lattice such as Mielke's flat-band ferromagnetism~\cite{Mielke1991a,Mielke1991b,Mielke1992}
or M\"uller-Hartmann ferromagnetism~\cite{Muller-Hartmann1995} and the presence of a van Hove singularity suggests the appearance of a Stoner instability at weak coupling~\cite{Hanisch1995,Honerkamp2003,Gao2007,Martin2008}.


Magnetic polarons play a crucial role for understanding the emergence of magnetic correlations in a wide range of materials~\cite{Alexandrov2010}. In seminal works by Nagaev~\cite{Nagaev1967,Nagaev1968,Nagaev1992},
it was proposed that conduction electrons doped onto antiferromagnetic semiconductors 
would generate significant ferromagnetic regions around them, the so-called ``ferrons'', wherein 
the electrons would become self-trapped. Similar phenomena were subsequently discussed for dilute doped magnetic semiconductors, such as Ga$_{1-x}$Mn$_x$As and Ge$_{1-x}$Mn$_x$. In these materials random potential constrains the motion of doped holes to regions of finite size and ferromagnetic bubbles form around each dopant.
With increasing dopant density, such bubbles begin to overlap, and ferromagnetic transition can be understood as a percolation transition of localized magnetic polarons~\cite{Kaminski2002}. 
Magnetic polarons have also been investigated within the Hubbard model~\cite{Bulaevskii1968,Brinkman1970,Trugman1988,Schmitt1988,Shraiman1988,Sachdev1989,Kane1989,Dagotto1989,Boninsegni1991,Boninsegni1992,White1997,White2001,Devereaux2015,Devereaux2015b,Grusdt2018,Grusdt2018,Koepsell2019,Soriano2020,Devereaux2020,Geoffrey2021,Koepsell2021}, and the emergence of ferromagnetic polarons, 
commonly referred to as Nagaoka's polarons, has also been examined in square lattices~\cite{White2001}. In this case Nagaoka polarons emerge only in the very-strong coupling regime $U/t\gtrsim 200$.
\begin{figure}[t!]
        \centering
        \includegraphics[width=1\columnwidth]{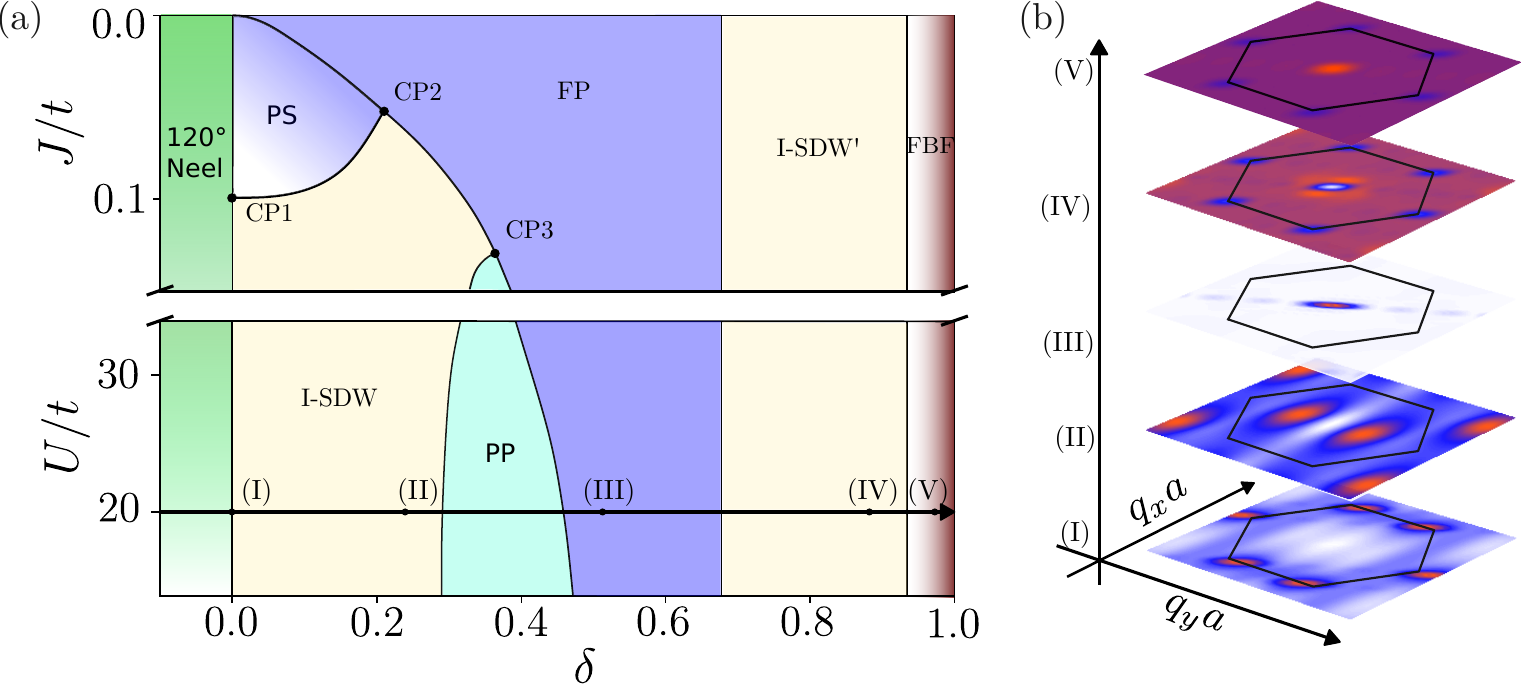}
        \caption{Schematic magnetic phase diagram of the Hubbard model in the triangular lattice in (a). The top (bottom) panel represents results obtained with the $t-J$ (Hubbard) model. We encounter a commensurate spin density wave (C-SDW) corresponding to 120$^{\circ}$ order (green regions), a phase separated (PS) regime (blue-to-white regions) where the system spatially phase separates in a ferromagnetic doublon-rich region and an undoped region with C-SDW order, a fully polarized (FP) state (full blue regions), a partially polarized (PP) state (light blue region), and a flat-band ferromagnet (FBF)  state (red regions). We encounter three critical points (CP1, CP2 and CP3) where different critical lines merge. In (b) we show the spin structure factors corresponding to the different phases. }
        \label{Fig:PhasDiagr}
\end{figure}

Recent experimental progress of quantum simulators made it possible to observe magnetic polarons directly. In quantum dot systems, Nagaoka-Thouless ferromagnetism 
has been observed in a small $2\times 2$ plaquette~\cite{Dehollain2020}. Experiments with cold atoms in triangular optical lattices have  demonstrated the 
tendency of the system towards ferromagnetism (antiferromagnetism) upon particle (hole) doping of the Mott insulating state~\cite{Greiner2022}. Additionally, two different cold atom experiments
have reported direct imaging of antiferromagnetic (Haerter-Shastry's) and ferromagnetic (Nagaoka's) polarons in the triangular lattice~\cite{Lebrat2023,Prichard2023}. These recent experimental advances provide strong motivation to explore theoretically the phase diagram of the doped Fermi Hubbard model. In this paper, we focus on fermion densities higher than 1, for which we find interesting competition between kinetic ferromagentism driven by charge carriers (doublons) and antiferromagnetic exchange interactions. While previous theoretical studies have explored the emergence of ferromagnetism in the triangular lattice with variational wavefunctions~\cite{Hanisch1995,Ogata2002,Honerkamp2003,Ogata2004,Giamarchi2006}, a full picture of the phase diagram is still unclear.

\textbf{Itinerant magnetism in the triangular lattice Hubbard model.} In this work we study the Hubbard model in the triangular lattice
\begin{align}
    \hat{H} &= -t \sum_{\langle i,j\rangle,\sigma } 
    \left(\hat{c}_{i,\sigma}^{\dagger}\hat{c}_{j,\sigma} + \textrm{h.c.}  \right)
     + U \sum_{i}\hat{n}_{i,\uparrow}\hat{n}_{i,\downarrow} ,\label{Eq:Hubb}
\end{align}
where $\hat{c}_{i,\sigma}^{\dagger}$ ($\hat{c}_{i,\sigma}$) creates (destroys) a
fermion with spin $\sigma$ at site $i$, 
$\hat{n}_{i,\sigma}=\hat{c}_{i,\sigma}^{\dagger}\hat{c}_{i,\sigma}$ 
corresponds to the particle number operator, $\langle i,j\rangle$ 
denotes nearest-neighbor sites, $t>0$ is the hopping strength and $U$ is the 
on-site interaction strength. 
\begin{figure}[t!]
        \centering
        \includegraphics[width=1\columnwidth]{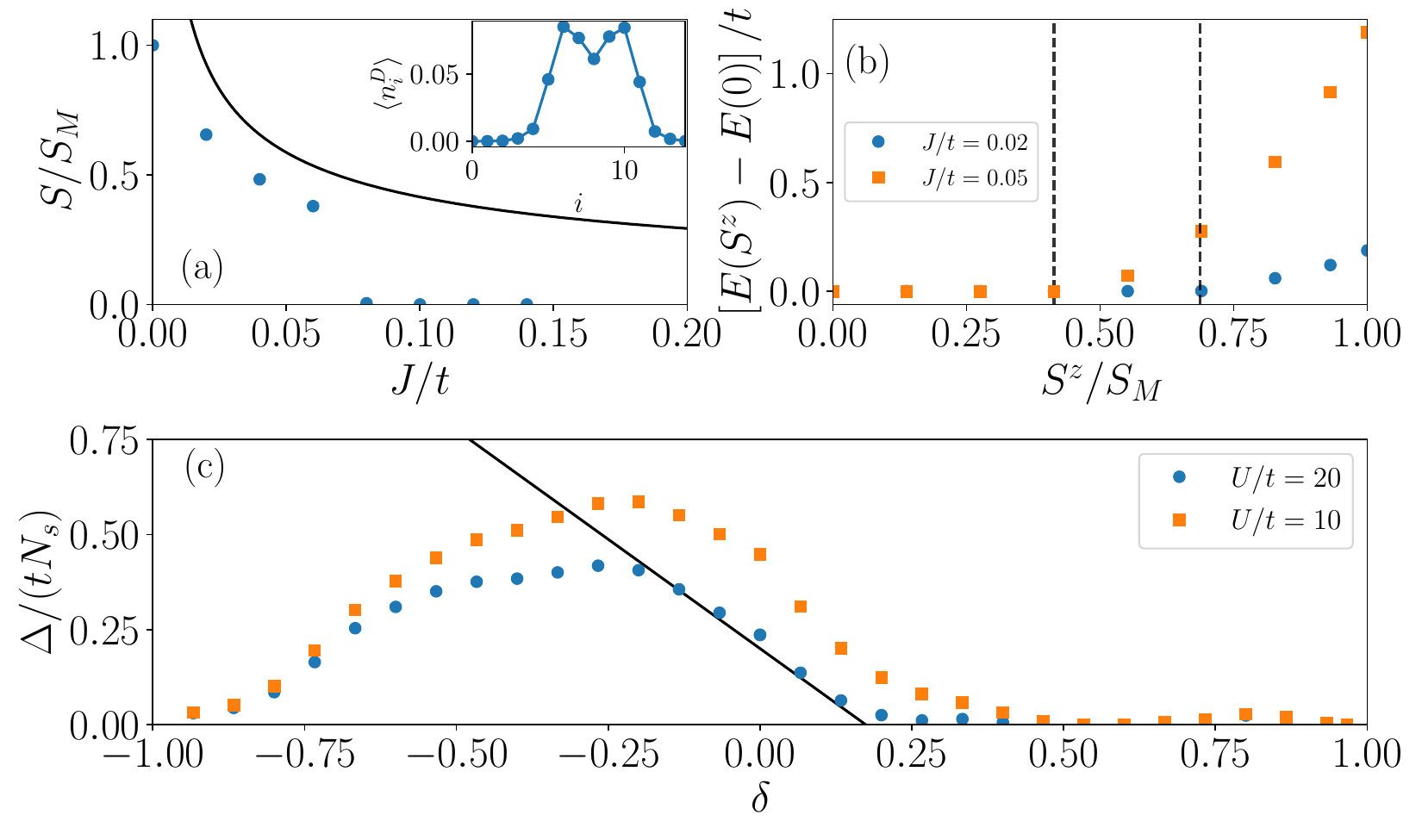}
        \caption{In (a) we represent the total spin $S$ as a function of the superexchange interaction $J/t$
        for a $15\times 4$ system with two doublons. Continuous black line corresponds to Eq.~\eqref{Eq:Spin}. In the inset panel we show the doublon density profile for $J/t=0.06$ and $N_D=2$. Two doublons bunch together in space denoting an effective attraction.
        In (b) we show the energy dependence with the total magnetization for two doublons and $J/t=1/20,1/50$. The dashed line denotes the total spin of the ground state for $S^z=0$.
        In (c) we show the polarization energy $\Delta$ as a function of the doping $\delta$
        for two different interaction strengths. The continuous black line denotes the linear dependence $\Delta/(tN_s)\sim 4t/U-1.15\delta$.}
        \label{Fig:tJ_Ferro}
\end{figure}

To obtain the ground state of Eq.~\eqref{Eq:Hubb} we perform density matrix renormalization group (DMRG) simulations implementing the U(1)$\times $U(1) symmetry in YC4 and YC6 cylinders~\cite{Szasz2020} with a maximum number of sites $N_s=60$ and a maximum bond dimension $\chi\sim 6000$. Moreover, to study the strongly interacting regime $U/t\gtrsim 40$ we perform simulations of the effective $t-J$ model with a superexchange interaction $J=4t^2/U$ and neglect three-site terms.

In Fig.~\ref{Fig:PhasDiagr} we present the magnetic phase diagram of the Hubbard model in the triangular lattice as a function of doping $\delta=n-1$ and on-site interaction $U/t$. For small hole doping ($\delta<0$) and strong interactions, we observed enhancement of the 120$^{\circ}$ commensurate spin density wave order (C-SDW). The mechanism of this enhancement has been discussed elsewhere~\cite{Morera2023}. For small doublon doping ($\delta>0$) and for $U/t>40$,  each doped doublon develops a Nagaoka-type ferromagnetic bubble around itself. In the ground state we do not find a homogeneous Nagaoka's polaron gas state, but phase separation between the doublon rich fully polarized state and a half-filled C-SDW state. In the phase separated regime, the system as a whole exhibits partial polarization and the total spin increases with the doublon density. For weaker interactions ($U/t<40$), we observe that a small doublon doping results in the formation of an incommensurate spin density wave (I-SDW). In this phase, peaks of the spin structure factor $\chi(q)$, which is defined in Eq.~\eqref{Eq:SQ}, are shifted from the $K$ points towards zero momentum by an amount that increases monotonically with increasing doublon density. We will refer to momenta corresponding to the peaks of the spin structure factor as $q^*$. We point out two similarities between the I-SDW phase and the stripe phase of cuprates: 1) when there are incommensurate peaks, we no longer see commensurate peaks at $K$; 2) for small doping we have linear scaling  $|q^*-K|$ with doping $\delta$ (similar to ``Yamada plot'' in cuprates~\cite{Yamada1998}). However, we do not observe sharp peaks in the density structure factor as expected for a stripe phase. Therefore, we conclude that the I-SDW state is a spiral phase with no clear density structure.
In Fig.~\ref{Fig:PhasDiagr} this phase corresponds to point (II).

At a critical doublon doping, we obtain a transition towards a fully polarized (FP) state for all values of $U/t$ considered, see (III) in Fig.~\ref{Fig:PhasDiagr}. Additionally, the critical doublon doping decreases with increasing on-site repulsion and it vanishes in the limit $U/t\rightarrow \infty$. A robust FP state appears in this limit for a wide range of doublon dopings $0<\delta\lesssim 0.7$ in our widest cylinder. Furthermore, we observe that ferromagnetism is centered around doping $\delta \sim 1/2$ in the intermediate interaction regime $U/t\sim20$ signaling the relevance of the van Hove singularity and the possible connection to a Stoner instability. The transition towards the FP state can be preceded by an I-SDW, a PS state or a partially polarized (PP) phase depending on the strength of the on-site interaction. The PP phase precedes the FP one by lowering the on-site interaction. It is characterized by a peak in the spin structure factor centered at zero momentum and a finite non-saturated magnetization. 

We identify three critical points (CPs) in the phase diagram in which two critical lines merge, see CP1, CP2 and CP3 in Fig.~\ref{Fig:PhasDiagr}. At infinitesimal doublon doping we obtain a CP (CP1) characterizing the transition from Nagaoka's polaron to Brinkman-Rice-Shraiman-Siggia type polaron~\cite{Brinkman1970,Shraiman1988,White2001}. At higher doublon dopings we encounter two different critical points (CP2 and CP3) which signal that the PP and PS states cannot be smoothly connected. Note that in each of the critical points CP2 and CP3 three different phases meet. We cannot rule out the possibility that with increasing width of the cylinders, critical points CP2 and CP3 start approaching each other and in the two-dimensional system we will have a single critical point in which four different phases meet. 

\begin{figure}
    \centering
    \includegraphics[width=1\columnwidth]{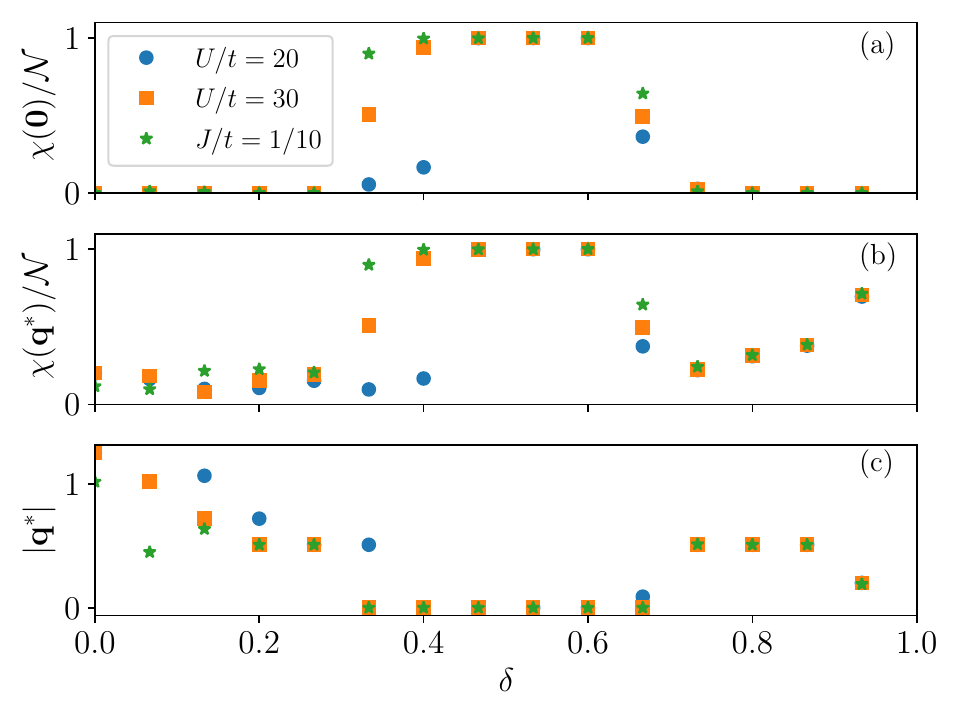}
    \caption{Normalized spin structure factor 
    $S(\mathbf{q})$ at null momentum in (a), at momentum  $\mathbf{q}^*$ in (b) and magnitude of the momentum $|\mathbf{q}^*|$ in (c) as a function of the doping factor $\delta$ for different interaction strengths computed in a YC4 cylinder. }
    \label{Fig:PeakSpin}
\end{figure}

Finally, in the large doublon doping regime ($0.8>\delta > 0.7$), we encounter an I-SDW' phase which presents dips at the $K$ points in the spin structure factor distinguishing it from the I-SDW phase at low doublon doping, see (IV) in Fig.~\ref{Fig:PhasDiagr}. Moreover, we observe that a ferromagnetic state develops in the limit of doping with two holes the band insulator at $\delta=1$, see (V) in Fig.~\ref{Fig:PhasDiagr}. We name this phase flat-band ferromagnetic (FBF) phase and we interpret it as M\"uller-Hartmann ferromagnetism~\cite{Muller-Hartmann1995}. It can also be understood as Mielke's flat band ferromagnetism~\cite{Mielke1991a,Mielke1991b,Mielke1992} in the limit of infinitesimal doping. We point out however, that we only find this regime for two holes regardless of the system size. Hence in the thermodynamic limit it corresponds to the vanishing hole density, see~\cite{Pieri1996} for a discussion of ferromagnetism stability at finite density. 

Having discussed the main features of the phase diagram, we now present the observables that we use to identify the different phases. One of the key characteristics that we study is the spin structure factor
\begin{align}
    \chi(\mathbf{q}) = \sum_{ij} \left(\langle \mathbf{S}_i \mathbf{S}_j \rangle - \langle \mathbf{S}_i\rangle\langle \mathbf{S}_j \rangle\right)e^{i \mathbf{q}\cdot \left(\mathbf{r}_i - \mathbf{r}_j \right)}.
    \label{Eq:SQ}
\end{align}
We note that the total spin of the system can be related to the same correlation function using the relation $\chi(\mathbf{0}) = S \left(S +1 \right)$. We use analysis of the total spin of the system to identify PP and FP phases. To confirm the nature of the PP phases we also compute ground state energy of the system for different values of the total magnetization $S^z$. Such magnetization-energy relation $E(S^z)$ has a simple physical interpretation. When the system has a tendency to develop ferromagentic order, fixing $S^z$ to a value that is smaller than the preferred magnetization, means that there will be additional magnetization in the perpendicular direction. Hence for a PP state we expect energy to be independent on magnetization until magnetization reaches the preferred total spin of the system. When $S^z$ increases beyond the preferred magnetization, energy will start increasing with increasing magnetization. Another experimentally relevant quantity that we analyze is the energy difference between the fully polarized state, $(S^z=S_M)$, and the unpolarized state with $(S^z=0)$, which denotes the energy cost to fully polarize the system. Finally, to detect the appearance of phase separation we compute the doublon density $\langle \hat{n}_i^D \rangle \equiv \langle \hat{n}_{i,\uparrow} \hat{n}_{i,\downarrow} \rangle$ and inspect where the density is localized in a finite region of the system.

\textbf{Magnetic polarons and effective interaction between them.} To understand the low-doping regime we first examine
the two-doublon problem. 
We focus on the regime $U/t\gg 1$, when the $t-J$ model
accurately describes the Hubbard model. Superexchange
interaction $J$ favors a 120$^0$ C-SDW configuration of spins which, based on our simulations,
leads to the ground state energy $E_J/N_s\sim -1.25 J$, where $N_s$ is the number of sites. 
However, a single doublon minimizes its constrained kinetic energy by inducing a ferromagnetic background around it as a result of the
constructive pattern of interference of different propagation paths. The competition
between these two terms leads to a Nagaoka's polaron of finite radius $R$. The doublon's
propagation is confined within the ferromagnetic region, and therefore, its kinetic energy can be approximated as $E_t/t\sim (2.4)^2/R^2$ in the continuum limit, as shown in~\cite{Nagaev1992,White2001}.
Considering both the kinetic and magnetic energy of the Nagaoka's polaron, we obtain the total spin $S = \pi R^2/2$,
\begin{align}
     S  \sim 1.95 N_D  \sqrt{t/J},
    \label{Eq:Spin}
\end{align}
where we assumed that each Nagaoka's polaron contributes to the total spin of the system in the same way. 
The size of the Nagaoka's polarons increases as the superexchange interaction is reduced and therefore, the total spin of the system increases. When polarons cover the entire system a long-range ferromagnetic state is formed in agreement with Nagaoka-Thouless' theorem.

To verify the formation of Nagaoka's polarons we numerically obtain the total spin as aforementioned.
As superexchange interaction, $J/t$, is decreased, the total spin $S$ starts to increase until it reaches its saturation value $S_M = (N_s-N_D)/2$ for $J/t\rightarrow 0$, see Fig.~\ref{Fig:tJ_Ferro} (a). Moreover, two polarons experience an effective interaction induced
by the magnetic environment. If two doublons share their ferromagnetic polaron bubbles, they can 
reduce the superexchange energy gain by reducing the area of the shared ferromagnetic bubble, but in doing so, they incur a higher kinetic energy due to the Pauli exclusion principle. 
In the regime $J/t<0.1$, two polarons have a tendency to come close to each other and share their
ferromagnetic bubbles. This signals an attractive polaron interaction, see Fig.~\ref{Fig:tJ_Ferro} (a).
However, for $J/t>0.1$, the total spin vanishes and two Nagaoka's polarons effectively repeal each other.

To confirm the partially polarized nature of the PS regime we compute its magnetization-energy relation $E(S^z)$, see Fig.~\ref{Fig:tJ_Ferro} panel (b). The ground state exhibits no energy dependence with the magnetization until a critical magnetization is reached at which the energy starts to increase. Moreover, the critical magnetization coincides with the total spin of the ground state at zero magnetization confirming that each Nagaoka's polaron creates a finite region of space that costs no energy to polarize in any direction. 

\textbf{Density-induced ferromagnetism and phase separation.} 
We now consider the regime of low doping, $\delta<0.1$, and large on-site interaction, $J/t<0.1$, see Fig.~\ref{Fig:PhasDiagr}. In this case we find phase separation between a doublon-rich ferromagnetic region
and an undoped region with 120$^o$ C-SDW order. 
In this regime, the total spin increases linearly with the number of doublons denoting
that the ferromagnetic region increases with doublon doping. Consequently the region with C-SDW order diminishes leading
to a state which exhibits partial polarization. 
By increasing the
doublon density the ferromagnetic puddle starts to increase, eventually occupying the entire system. At this point, doublons delocalize throughout the entire system and 
the system transitions to a long-range ferromagnetic state. 

In the intermediate interacting regime ($40>U/t>20$) we perform simulations of the Hubbard model Eq.~\eqref{Eq:Hubb} in YC4 cylinders. 
We do not observe signatures of a PS state at $U/t\leq 40$ for small doublon dopings. Instead, we note that doublons establish an I-SDW at finite momenta, see Fig.~\ref{Fig:PhasDiagr} (II) and the associated momenta decrease with increasing doublon doping, see Fig.~\ref{Fig:PeakSpin}. We observe that the momenta suddenly jump from a finite to a vanishing value at a critical doublon doping, denoting a first order phase transition between the I-SDW and the FP phases. In Fig.~\ref{Fig:tJ_Ferro} (c) we present the polarization energy for the Hubbard model at different interaction strengths. 
The polarization energy linearly decreases (increases) upon particle (hole) doping,
signaling that dopants favor ferromagnetism or antiferromagnetism in the particle or hole doping regimes, respectively. Furthermore, the linear relation 
suggests that each dopant contributes to the magnetic order, reinforcing the idea that doublons trigger the formation of ferromagnetic regions. 
When polarization energy vanishes, a transition towards a long-range ferromagnet occurs.


\textbf{Stability of long-range ferromagnetism at large doublon dopings.} 
In the doping regime $\delta \sim 0.4-0.6$
we observe a strong ferromagnetic phase in both YC4 and YC6 cylinders, that persists up to the intermediate interacting regime $U/t\sim 20$, see Fig.~\ref{Fig:PeakSpin}. 
The ferromagnetic phase in the intermediate interacting regime is centered around $\delta\sim 1/2$ signaling the importance
of the van Hove singularity and suggesting that the mechanism leading to a ferromagnetic
state is likely to be a Stoner instability. We note that there is no global
distinction between Stoner-type ferromagnetism and the Nagaoka-type kinetic ferromagnetism aforementioned, 
since they correspond to the same phase. Notably, our simulations in YC6 cylinders reveal
that these two mechanisms can be continuously connected by increasing $U/t$. 

Long-range ferromagnetism becomes unstable
around $1>\delta \gtrsim 0.7$ in both geometries (YC4 and YC6) even in the limit $U/t\rightarrow \infty$. 
In this regime, we note that the spin structure factor exhibits minima at the $K$, $K'$ and $\Gamma$ points and an I-SDW' is formed, see Fig.~\ref{Fig:PhasDiagr} (IV). Moreover, we observe that the polarization energy per site becomes very small in the I-SDW' phase, see Fig.~\ref{Fig:tJ_Ferro} (c).
Furthermore, we observe that the spin structure factor becomes almost independent on the interaction strength $U/t$ in this high doping regime, see Fig.\ref{Fig:PeakSpin}. 

Finally, let us discuss the 
vanishing doping situation. Specifically, we explore the case of two holes introduced on top of the band insulator $\delta \rightarrow 1^{-}$. By performing a low-energy expansion around
the $K$ and $K'$ points we obtain a ground state which is a spin triplet and a valley (orbital) singlet, see~\cite{Muller-Hartmann1995,Chengshu2023} for a similar discussion. In this limit, 
the ground state exhibits maximum spin and the spin structure factor exhibits features at the $K$, $K'$ and $\Gamma$ points, see Fig.~\ref{Fig:PhasDiagr} (V),
reflecting that the holes populate the $K$ and $K'$ valleys in an antisymmetric way. This has to be contrasted with the aforementioned long-range ferromagnetism, which is characterized by a single peak in the spin structure factor at the $\Gamma$ point. Moreover, flat-band ferromagnetism
survives for any interaction strength $U/t$. Therefore, our results suggest that M\"uller-Hartmann ferromagentism appears 
in the vanishing doping limit ($\delta\rightarrow 1^-$), but it becomes unstable at finite doping and it is not connected
with the ferromagentic phase at intermediate dopings ($0.7>\delta>0$) even in the limit $U/t\rightarrow \infty$. Although we cannot entirely rule out the possibility of finite density M\"uller-Hartmann ferromagnetism being present in the full 2D limit, 
our simulations conducted in our widest cylinder do not support its existence.

\textbf{Conclusions and outlook.} We have investigated the ground state magnetic properties
of the Hubbard model in the triangular lattice at finite doping. In the low doublon doped regime and for $U/t>40$ we find that individual doublons form Nagaoka's ferromagentic polarons. Attractive interaction between such polarons results in a phase separation between a half-filled state and a doublon rich phase with full polarization. However, for $U/t<40$ we observe an incommensurate spin density wave. By increasing particle doping we find a transition towards a long-range ferromagnetism for all interaction strengths considered. 
Therefore, our results show that long-range ferromagnetism is stable in the particle doped triangular lattice Hubbard model for a wide range of parameters.

In the intermediate interacting regime $U/t\sim 10$ one also expects the emergence of a chiral spin liquid~\cite{Szasz2020,Chen2022}. 
However, the fate of the chiral spin liquid upon particle doping is still unclear and we are currently 
limited by system size to speculate on the chirality of the ground state. Nevertheless,
recent works have pointed out that a chiral metallic phase arises when the spin liquid is hole doped~\cite{Zhu2022}. Consequently,
a comprehensive exploration of the doublon-doped system in this intermediate interacting regime will
provide deeper insights on the chiral nature of the ground state. Furthermore, hole spectral functions have been shown to be a useful observable that reveals the chiral spin nature of the ground state and the appearance of dynamical Nagaoka's polarons~\cite{Kadow2022,Kadow2023}. In combination with the recent photoexcitation probes discussed in~\cite{Morera2023_photo} the nature of the Nagaoka's polarons could be explored in current cold atoms experiments.

Finally, our results reveal the occurrence of phase separation within the low doublon-doped regime, 
signaling the presence of an effective attraction between Nagaoka's polarons. However, phase separation can be suppressed upon inclusion of non-local Coulomb interaction, which can be relevant for experiments with TMDCs. We envision that Coulomb interaction
could stabilize a bipolaron without leading to phase separation which could pave the way to explain the emergence of 
superconductivity in doped triangular Mott insulators. Furthermore, thermal fluctuations can also overcome attraction between Nagaoka's polarons and lead to a uniform polaronic phase.

\textbf{Note added.} When this paper was being finalized, we learned about paper~\cite{Parameswaran2023} that addressed the phase diagram for the particular filling factor $\delta=1/2$.

\textbf{Acknowledgments.} We acknowledge useful discussions with Waseem S. Bakr, Utso Bhattacharya, Annabelle Bohrdt, Anant Kale, Youqi Gang, Markus Greiner, Fabian Grusdt, Lev Haldar Kendrick, Wen-Wei Ho, Atac Imamoglu, Martin Lebrat, Max L. Prichard, Benjamin M. Spar, Muqing Xu, Zoe Z. Yan. 
We acknowledge support from the ARO grant number W911NF-20-1-0163, the SNSF project 200021\_212899, and the Swiss State Secretariat for Education, Research and Innovation (SERI) under contract number UeM019-1.
Tensor network
computations have been performed
using TeNPy~\cite{10.21468/SciPostPhysLectNotes.5}.

\bibliography{paper}

\end{document}